\documentclass[pra,aps,twocolumn,showpacs]{revtex4}

\begin{document}

\title{Measurement does not always aid state discrimination}
\author{Kieran Hunter}
\affiliation{Department of Physics, University of
Strathclyde, Glasgow G4 0NG, Scotland}

%\date{\today}

\begin{abstract}

We have investigated the problem of discriminating between nonorthogonal quantum
states with least probability of error. We have determined that the best
strategy for some sets of states is to make no measurement at all, and simply to
always assign the most commonly occurring state. Conditions which describe such sets of
states have been derived.

\end{abstract}


\maketitle

Finding the best measurement strategy to perform on some quantum
signal is an important problem, with particular relevance to quantum communication and computation.
The optimal measurement strategy will depend on the purpose of the measurement.
To optimise our strategy we must define a figure of merit function which provides a measure of the
appropriateness of a strategy. Then we must find the measurement which maximises (or minimises) the
chosen figure of merit.

A commonly considered example of this is the probability of incorrectly
identifying the state
$P_{e}$.
A set of necessary and sufficient conditions for a measurement to minimise $P_{e}$ (or more
generally the mean Bayes cost) are known \cite{Helstrom, Holevo, Yuen}, but a
solution to these conditions has only
been found in a small number of cases. These are: when there are only two possible states of
the signal \cite{Helstrom}, when the states are linearly independent
\cite{Ken73},  when
the states are equiprobable and sum to the identity \cite{Yuen}, the cases of
equiprobable 
symmetric \cite{Symerr} (or geometrically uniform \cite{Eldar})
and multiply symmetric \cite{Barsym}
states, and the case of three mirror symmetric
qubit states \cite{Erika1}.

In this paper we present an additional solution to these minimum error
conditions. We describe sets of states
where no measurement discriminates between the states better than assigning the a priori most
likely state to the signal.

To find an optimal strategy we describe the
measurement by its Probability Operator Measure (POM)
 elements $\hat{\Pi}_{k}$. These POM elements
 are operators which represent the probability of occurrence of each possible
 outcome of a measurement. The probability $P(k|j)$ of the outcome $k$ occurring
 given that the system was in its $j^{th}$ state $\hat{\rho}_{j}$ is
\begin{equation}
P(k|j) = \mathrm{Tr}(\hat{\Pi}_{k}\hat{\rho}_{j}). \label{probpi}
\end{equation}

For the POM elements  $\hat{\Pi}_{k}$ to represent probabilities, they must be
subject to
the following constraints:
\begin{enumerate}
\item All the $\hat{\Pi}_{k}$'s are Hermitian,
\item Their eigenvalues non-negative,
\item The total probability of all outcomes for any input sums to 1:
\begin{equation}
\sum_{k} \hat{\Pi}_{k} = \hat{1}.\label{eq:POMcond2}
\end{equation}
\end{enumerate}

The conditions for a measurement strategy $\{\hat{\Pi}_{k}\}$ to minimise the probability
of erroneously identifying the state of the signal can be easily derived from those for minimising the
mean Bayes cost \cite{Helstrom}
by an choosing the cost of being wrong to be a constant. The minimum error
conditions can then be stated as
\begin{equation}
\sum_{j} \hat{\Pi}_{j} p_{j} \hat{\rho}_{j} - p_{k} \hat{\rho}_{k} \ \geq \ 0 \ \forall \ k, \label{conG}
\end{equation}
which means that the operator on the left is both Hermitian and positive semidefinite. From
this can be derived the necessary condition \cite{Helstrom}
\begin{equation}
\left(\sum_{j} \hat{\Pi}_{j} p_{j} \hat{\rho}_{j} - p_{k} \hat{\rho}_{k}\right) \hat{\Pi}_{k} = 0 \ \forall \ k, \label{conE}
\end{equation}
which implies the Hermiticity of (\ref{conG}).

The derived condition (\ref{conE}) has an interesting property. It is always satisfied by the POM
\begin{equation}
\hat{\Pi}_{j} = \hat{1}, \hat{\Pi}_{k \neq j} = \hat{0}. \label{nomeasure}
\end{equation}
This POM corresponds to not making any measurement on the signal at all, and simply always
assigning the state $\hat{\rho}_{j}$ to the signal. In effect this POM is a consistent guessing
strategy.
This POM automatically makes the left side of (\ref{conG}) Hermitian.

When such a POM is used, the condition (\ref{conG}) becomes
\begin{equation}
p_{j} \hat{\rho}_{j} - p_{k} \hat{\rho}_{k} \ \geq \ 0 \ \forall \ k, \label{conNM}
\end{equation}
for some value of $j$.
When this condition is satisfied by the set of possible states of the signal $\{\hat{\rho}_{k}\}$
with prior probabilities $p_{k}$, there exists no measurement which distinguishes the states
better than guessing. We must now determine the physical meaning of (\ref{conNM}).

It is obvious that the state $\hat{\rho}_{j}$ must be the most likely state. This can be
verified be taking the trace of (\ref{conNM}), which gives $p_{j} \ \geq \ p_{k} \ \forall \ k$. It
is also clear that $\hat{\rho}_{j}$ cannot be a pure state (except in the trivial case where
all of the states are identical) since then the operator on the left of (\ref{conNM}) would
have a negative expectation value for some states. Indeed this tells us that it is not sufficient
for the state $\hat{\rho}_{j}$ to be mixed: this state must span the entire space of the states
$\hat{\rho}_{k}$. These are necessary, but not sufficient for (\ref{conNM}) to be satisfied.

Denoting the most likely state as $\hat{\rho}_{0}$, we can restate the condition
(\ref{conNM}) 
as the vector
inequality  ``In every direction, the
component  of the operator $p_{0} \hat{\rho}_{0}$ must be
greater than or equal
to the components of each of the operators $p_{k} \hat{\rho}_{k}$ in that direction.'':
\begin{equation}
\langle\phi|\left(p_{0} \hat{\rho}_{0} - p_{k} \hat{\rho}_{k}\right)|\phi\rangle \ \geq \ 0 \ \forall \
k,|\phi\rangle.
\end{equation}

If all of the states other than the most likely state $\hat{\rho}_{0}$  are pure
states, $\hat{\rho}_{k}=
|\psi_{k}\rangle\langle\psi_{k}|$, we can simplify this condition by noting
that the only significant vector $|\phi\rangle$ in this case is $|\phi\rangle=|\psi_{k}\rangle$.
The condition then reduces to
\begin{equation}
\langle\psi_{k}|p_{0} \hat{\rho}_{0} |\psi_{k}\rangle \ \geq \ p_{k} \ \forall \ k.
\end{equation}
If instead the $\hat{\rho}_{k}$ are mixed we shall obtain one such relation for
each pure state that $\hat{\rho}_{k}$ can be decomposed into, with $p_{k}$ being
multiplied by the weight of that pure state in $\hat{\rho}_{k}$.

This simplifies even further if the most likely state $\hat{\rho}_{0}$ is a
no-information (maximally
mixed) state,
$\hat{\rho}_{0}=\frac{1}{D}\hat{1}$ where $D$ is the dimension of the state space:
\begin{equation}
\frac{p_{0}}{D} \ \geq \ p_{k} \ \forall \ k. \label{simple}
\end{equation}

The best way to illustrate the significance of these states is by an example.

Let us consider a communication channel in which the signal can be any one of $N$ pure states
$|\psi_{k}\rangle$. The preparation of each of the $N$ states are equally likely, but there is
 also a chance that the preparation will fail and a completely random state will be sent.
Here we are viewing the preparation to be both the transmitter and the channel
itself, as it does not
matter where these failures occur. We wish to identify with least probability of error what was sent:
either a specific signal state or a failed transmission. Is there any point in measuring the received
signal?

In this example the only state which could satisfy the requirement to span the space is the `failed
transmission'. Since the signal is completely random in this case the only state which can be assigned
to the signal is $\hat{\rho}_{0}=\frac{1}{D}\hat{1}$, where $D$ is the dimension of the
space.

Since the $N$ possible signals are equally likely, we can set their probabilities
$p_{k}=p=\frac{1-p_{0}}{N}$. The simplified condition for the case of
discriminating unlikely pure
states from a single maximally mixed state (\ref{simple}) then gives
\begin{equation}
\frac{p_{0}}{D} \ \geq \ \frac{1-p_{0}}{N},
\end{equation}
which implies
\begin{equation}
p_{0} \ \geq \ \frac{D}{D+N}.
\end{equation}
If this inequality holds then there is no measurement which will distinguish the
signal states from a failed transmission.

For the case of three qubit signal states this would give $p_{0}\geq\frac{2}{5}$. Even at a
failure rate of only two fifths for these signal states, it is still impossible
to find any measurement which discriminates the
signal states and failure with less probability of error than always assuming that the preparation has
failed.

At this point it is worth discussing what this result means, and its limitations. The conditions on the
set of states such that the no measurement POM is optimal have a clear interpretation in terms of the
likelihood of the states. To understand this we must look at how the measurement affects the assignment
of the signal state.

Before we measure the state of the signal the only information we have about that state are the
prior probabilities of each possible state. Thus we assign these prior probabilities as the likelihood
of detecting each state, with the highest probability belonging to the most likely state. Once we have
measured the state we also know the measurement outcome. We can use Bayes rule with (\ref{probpi}) and
$p_{j}$ to calculate the probability $P(j|k)$ that the signal state was $\hat{\rho}_{j}$  given that the
measurement outcome was k. These $P(j|k)$ are the probabilities we assign to each state on the basis of
our updated information which now includes the result of our measurement.

If the a priori most likely state will remain the most likely state {\em whatever the result of any
measurement made}, then no measurement discriminates between the states better than guessing. This would
not be a surprising result if it were not for the fact that it is quite easy to obtain such a set of
states. It can also hold for some classical systems, if there is a severe restriction on the form of
the
measurements which can be made. Only using quantum systems and measurements can we say that it can be
satisfied for all physically possible measurements.

Just because we have established that we cannot identify the most likely state by a measurement does not
mean that there is no point in performing a measurement. One could, for example, try to identify the
next most likely state. In the communication example given earlier, we would forget
about trying to determine if the signal was real or a failed transmission and instead ask ``If we assume
that this signal is not a failed transmission, what is the most likely state of the signal?'' This will
determine which of the $N$ signal states is the most likely to have been
transmitted, but it would still be more likely that transmission failed.

We can also give up on any positive identification of the state and instead try to obtain as much
information as possible about the signal. The appropriate figure of merit for this would be the mutual
information gain $I$ \cite{Davies, Sasaki, Mizuno}
 which is always positive for any actual measurement and zero for our no measurement POM. That this
 gives a different result should not be surprising as the maximum information strategy is different from
 the minimum error strategy even for very simple examples \cite{Clarke1}.

In conclusion, we have found an interesting solution to the problem of discriminating between the
possible states of a quantum signal or system with least probability of error. 
For some sets of states it is possible to
satisfy the necessary and
sufficient conditions for minimum error by not making any measurement at all,
and simply assigning the most common state to the system. We have explored the
restrictions on such sets of states, and developed simplifications for these when the most likely
state is maximally mixed and also when the other states are pure states.

These results were illustrated by a quantum communications example, and can be
easily interpreted in terms of Bayes rule.

This work was supported by the UK Engineering and Physical Sciences Research Council.
The author would like to thank Professor Stephen M. Barnett for his helpful
advice on the production of this paper.


\begin{thebibliography}{99}

\bibitem{Helstrom} C.W. Helstrom, {\em Quantum Detection and Estimation theory}
(New York: Academic 1976).

\bibitem{Holevo}  A.S. Holevo, {\em Probabilistic and Statistical Aspects of Quantum
Theory} (Amsterdam: North-Holland 1982).

\bibitem{Yuen} H.P. Yuen, R.S. Kennedy, M. Lax,  IEEE Trans. Inf.
Theory {\bfseries IT-21} 125 (1975).

\bibitem{Ken73} R.S. Kennedy, M.I.T. Res. Lab. Electron. Quart. Progr.
Rep.
{\bfseries 110} 142 (1973).

\bibitem{Symerr} M. Ban, K. Kurokawa, R. Momose, O. Hirota,   Int. J. Theor.
Phys. {\bfseries 36} 1269 (1997).

\bibitem{Eldar} Y.C. Eldar, G.D. Forney Jr,   IEEE Trans. Inf. Theory
{\bfseries 47} 858 (2001).

\bibitem{Barsym} S.M. Barnett,  Phys. Rev. A {\bfseries 64} 030303(R)
(2001).

\bibitem{Erika1}  E. Andersson,  S.M. Barnett,  C.R. Gilson,  K. Hunter, 
 Phys. Rev. A. {\bfseries 65}, 052308 (2002).

\bibitem{Davies} E.B. Davies,   IEEE Trans. Inf. Theory {\bfseries
IT-24} 596 (1978). 

\bibitem{Sasaki}  M. Sasaki,  S.M. Barnett,  R. Jozsa,  M. Osaki and O. Hirota, 
 Phys. Rev. A. {\bfseries 59} 3325 (1999).

\bibitem{Mizuno} J. Mizuno, M. Fujiwara, M. Akiba, T. Kawanishi, S.M. Barnett,
 M. Sasaki, 
 Phys. Rev. A. {\bfseries 65}, 012315 (2002).

\bibitem{Clarke1}  R.B.M. Clarke,  V.M. Kendon,  A. Chefles, S.M. Barnett,
 E. Riis,  M. Sasaki,  
 Phys. Rev. A.  {\bfseries 64} 012303 (2001).

\end{thebibliography}
\end{document}